# Sub-mW 30GHz Variable-Gain LNA in 22nm FDSOI CMOS for Low-Power Tapered mm-Wave 5G/6G Phased-Array Receivers

Michele Spasaro and Domenico Zito

Dept. of Electrical and Computer Engineering, Aarhus University, Denmark

domenico.zito@ece.au.dk

*Abstract*—Next-generation cellular systems require low-power mm-wave phased-array ICs. Variable-gain LNAs (VG-LNAs) are key blocks enabling reduced hardware complexity, performance improvement and added functionalities. This paper reports a low-power 30GHz VG-LNA for mm-wave 5G/6G phased-array ICs, with a gain control of 8 dB for 18dB Taylor taper in a 30GHz 8x8 antenna array. The VG-LNA exhibits a peak gain of 16 dB in the high-gain state, consumes less than 1 mW and occupies an area of 0.20×0.22 mm$^2$.

*Keywords*—Gain control, low-noise amplifier, phased arrays, 5G/6G mobile communication.

## I. Introduction

The introduction of operating bands in the 5G new radio (NR) Frequency Range 2 (FR2), i.e., 24.25-52.6 GHz, and the progressive standardization of the next-generation cellular system is fostering the research on phased-array ICs for 5G wireless communications and beyond. The first-reported 28GHz IC for 5G [1] features 32 TX/RX (TRX) elements, each including a TX/RX switch, a 12dB gain LNA with NF of 6 dB including the losses of the TX/RX switch and off-mode PA, and a phase-invariant VGA enabling beam shaping, as in Fig. 1(a). Some applications can benefit of switchless TRX architectures, as in Fig. 1(b), typically asymmetrical for accomplishing communication needs and hardware efficiency. The absence of the lossy TX/RX switch allows relaxing the gain and noise requirements for the LNAs, enabling low-power designs. Moreover, variable-gain LNAs (VG-LNAs) can replace the VGA in the RX elements so to allow further reduction of the overall power consumption ($P_C$). VG-LNAs are also employed in front-end elements for phased-array ICs with digital beamforming [2]. Thereby, compact low-power VG-LNAs are key to the implementation of highly efficient phased-array ICs for future generation wireless transceivers.

Prior works [2-6] reported low-power VG-LNAs for 5G and 6G phased-array ICs. Reported gain control techniques include current steering [2, 3]; tunable resistive loads and bias current control [4]; bias current control with phase compensation [5]; and forward body bias [6]. Despite these techniques do not necessarily require further $P_C$ and area on chip, prior-art VG-LNAs [2-6] consume from a few units/tens of milliwatts and occupy silicon areas larger than 0.1 mm$^2$, owing to lossy and large passive components such as spiral inductors, transformers, transmission lines (TLs), and differential topologies [4-5]. No works to date have reported sub-mW mm-wave VG-LNAs.

This paper presents a sub-mW 30GHz VG-LNA in 22nm FDSOI CMOS for low-power tapered phased-array receivers.

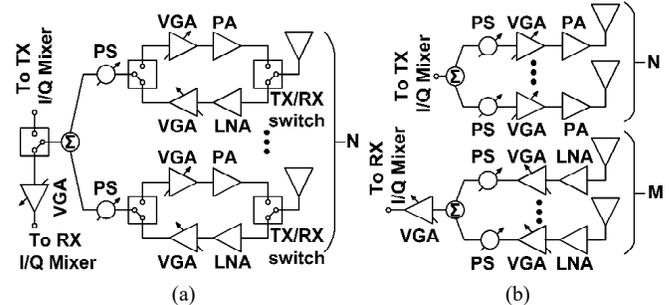

Fig. 1. Phased array TX/RX architectures for phased array ICs with RF phase shifting: (a) Symmetric architecture with TX/RX switch; (b) symmetric/asymmetric architecture without TX/RX switch.

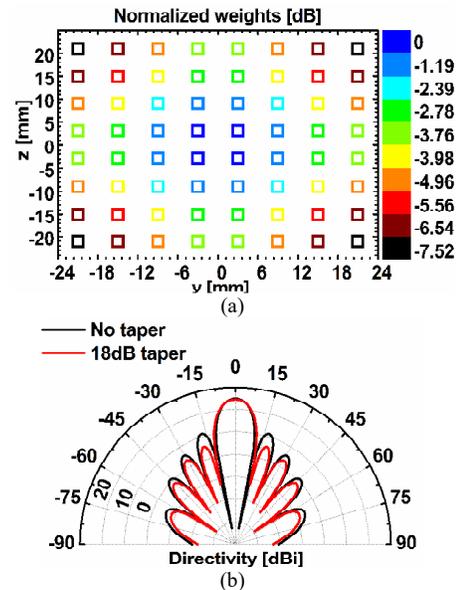

Fig. 2. Gain control and beamforming in 30GHz 8×8 3GPP TR 38.901 compliant-antenna array (6mm pitch) with 18dB Taylor window taper. (a) Normalized weights. (b) Radiation pattern of the antenna array.

The paper is organized as follows. Section II addresses the gain-control requirements for the VG-LNA. Section III summarizes the main features of the VG-LNA with special emphasis on gain-control performances. Section IV reports the experimental results. Section V draws the conclusions.

## II. Requirements for Gain Control

The VG-LNA has been designed to implement a Taylor window taper up to 18dB with three approximately equal side lobes in a planar 8×8 antenna array operating at 30 GHz. The VG-LNA can also be employed in untapered phased arrays,

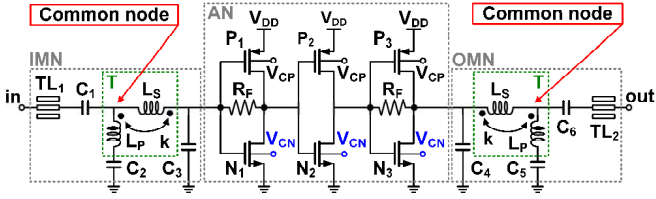

Fig. 3. Low-power VG-LNA circuit. Dashed boxes identify the custom-designed transformer (T) with its common node, active network (AN), input matching network (IMN) and output matching network (OMN).

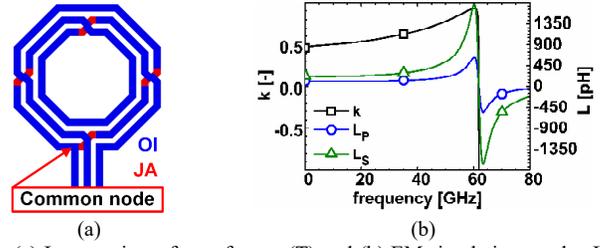

Fig. 4. (a) Layout view of transformer (T) and (b) EM simulation results. $L_P$, $L_S$, and k amount to 119 pH, 267 pH, and 0.59, respectively, at 30 GHz.

which take advantage of the gain control to improve the dynamic range of the receiver.

Fig. 2 reports the normalized weights and the directivity of the antenna array, with and without taper, simulated in the MATLAB phased-array system toolbox, considering the built-in model of an antenna element compliant with 5G NR technical report 3GPP TR 38.901 and a pitch of 6 mm. The results show that side lobe level of -18dB can be achieved with 7.5dB gain control range, which is aligned with experimental results in [1]. The VG-LNA is designed to exhibit gain control range of 8 dB with sub-mW $P_C$ for all gain states, identified by the measured gain at 30 GHz, and with peak gain of 16 dB in the high-gain state, so accomplishing the low-power tapered array requirements. A noise figure of 6 dB was considered, similarly to other designs, e.g., [1]. In absence of blockers, the array is operated in condition of maximum sensitivity, i.e., uniform weights for maximum directivity, and all the VG-LNA elements are in the high-gain state. In this condition, the signal-to-noise ratio after the beamforming network is improved by M = 64, i.e., 18 dB, with respect to a single front-end element [7].

### III. VG-LNA FEATURES

Fig. 3 reports the VG-LNA circuit. The active network (AN) exploits complementary current reuse and consists of three gain stages to achieve a nominal gain of 12 dB with a $P_C$ of about 0.6 mW. The back-gate voltages $V_{CP}$ and $V_{CN}$ are provided through the metal grid connected to the DC pads. The gain is varied by applying a single control voltage $V_{CN} = V_C$ to the back-gates of all n-FETs. $V_{CP}$ is set equal to 0 V.

Unlike the common microwave/mm-wave design approach, the amplifier stages of the AN do not include spiral inductors/transformers for peaking and inter-stage matching networks (MNs), which require connecting the transistors to the top metal layers, used for the passives all through the back-end-of-line (BEOL) and then introducing substantial losses. This design paradigm shift leads to very compact layouts with reduced parasitics and allows taking better advantage of ultra-scaled MOSFETs' performances, as these are not reduced by lossy interconnects, whose effects are predominant in low-power designs at high frequencies. However, the effectiveness of this approach reduces as the operating frequency increases, due to the inherent parasitic capacitance of the transistors.

All transistors have gate width (W) of 3.9 μm and $V_{DD}$ amounts to 800 mV. The AN exhibits an input impedance of 290-j450 Ω at 30 GHz in the 12dB gain state. This input impedance is transformed to 50 Ω through the input matching network (IMN), featuring the cascade of two L-networks: $C_1$-$L_P$ and $L_S$-$C_3$, with the magnetic coupling for boosting the effective Q-factor of the two spiral inductors and reduce the footprint on chip. $L_S$-$C_3$ transforms the input impedance of AN in an intermediate impedance $Z_{IM}$ that is used as a free parameter to enable the integrated implementation of the IMN and, to a limited extent, to optimize the noise performance of the VG-LNA [8]. The coupling factor k is an additional design parameter employed for the minimization of the cascade noise [8]. Similar considerations apply to the output matching network (OMN), in which the gain performance is optimized instead. Fig. 4 illustrates the layout of the coupled inductors, i.e., transformer T, implemented with the topmost copper metal layers (JA and OI), and reports the results of the EM simulations. Each MN includes a short (50 μm) 50Ω coplanar waveguide to the pads. The capacitors of the MNs are sized as follows: $C_1$ = 238 fF, $C_2$ = 548 fF, $C_3$ = $C_4$ = 26 fF, $C_5$ = 730 fF, $C_6$ = 550 fF. The AN includes two resistors ($R_F$) of 10 kΩ.

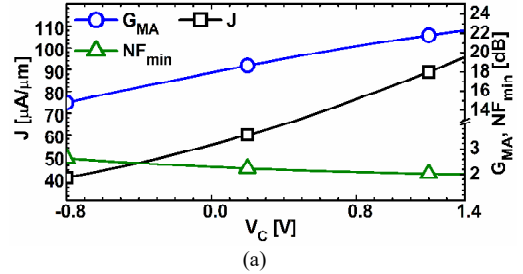

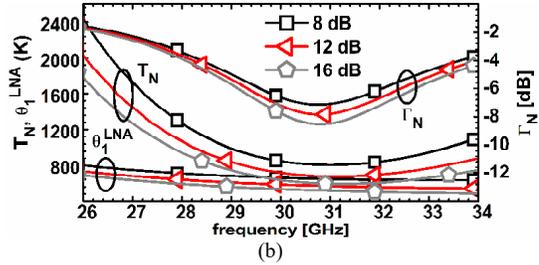

Fig. 5. PLS results: (a) Transistor J, and $G_{MA}$ and $NF_{min}$ of the AN at 30 GHz, vs. $V_C$; (b) equivalent noise temperature ($T_N$), $\theta_1^{LNA}$ ($T_N$ theoretical lower bound), and impedance-matching coefficient ($\Gamma_N$) between IMN output impedance and AN optimum-noise impedance [8] of the VG-LNA vs. frequency, for low, medium (nominal) and high gain states.

Fig. 5 reports post-layout simulation (PLS) results. In Fig. 5(a), the maximum available gain ($G_{MA}$) and minimum NF ($NF_{min}$) of the AN at 30 GHz, and transistor current density (J) are plotted as a function of the control voltage $V_C$. J increases from 41 to 94 μA/μm as $V_C$ varies from -0.75 to 1.35 V, leading to an increase of the $G_{MA}$ from 15 to 22 dB, and a reduction of the $NF_{min}$ from 2.6 to 2.0 dB as J approaches the optimum-noise current density of about 0.1-0.15 mA/μm.

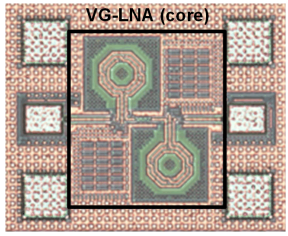

Fig. 6. Die micrograph. Core area: 0.20×0.22 mm².

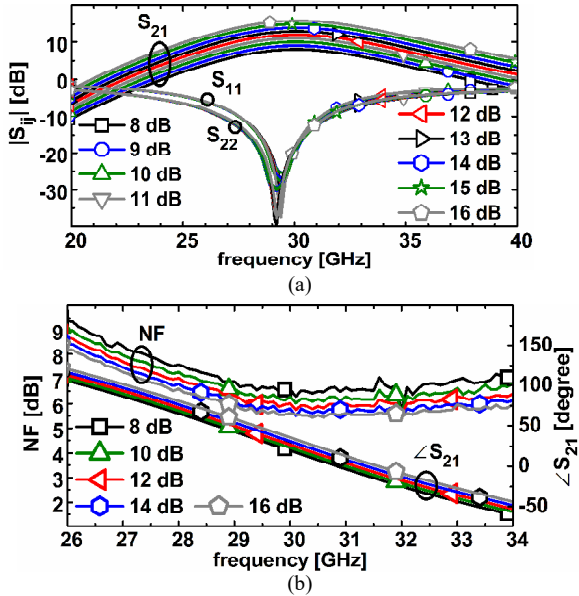

Fig. 7. Measured (a) S-parameters; (b) ∠$S_{21}$ and NF of the VG-LNA, for all gain states.

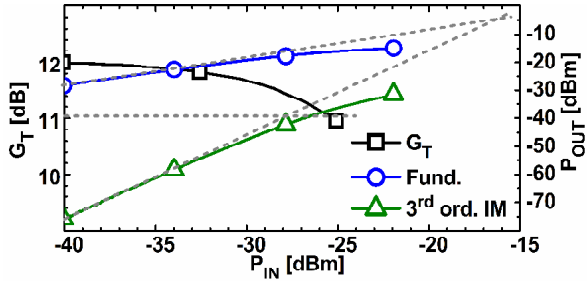

Fig. 8. Measured linearity performances of the VG-LNA in the nominal (12dB) gain state: Transducer gain ($G_T$) at 30 GHz and average power of the fundamentals and 3rd order intermodulation products (IM) for input tones at 30 and 30.05 GHz.

## IV. EXPERIMENTAL RESULTS

Fig. 6 reports the die micrograph. The VG-LNA have core area of 0.20×0.22 mm² and silicon area of 0.32×0.26 mm² including RF pads. The VG-LNA was measured on die with PNA-X N5245A. Power calibrations were carried out with the power meter N1914A, power sensor N8487A, and ECal N4693A. Fig. 7 reports the S-parameters $S_{21}$, $|S_{11}|$ and $|S_{22}|$, and NF for all gain settings. In the high-gain state, the VG-LNA exhibits a peak gain of 16 dB at 30.1 GHz and an NF of 5.5 dB with a $P_C$ of 0.97 mW. In the low-gain state, the $P_C$ amounts to 0.41 mW. The $S_{12}$ is lower than -33 dB for all gain states. Fig. 8 and Table 1 report the measured input-referred $P_{1dB}$ ($IP_{1dB}$) and $IP_3$ ($IIP_3$) of the VG-LNA for all gain settings. Two-tone tests were carried out with two tones at $f_1$=30 GHz and $f_2$=30 GHz+Δf, with Δf=50, 100, 200, and 400 MHz, as per 5G channelization. Fig. 8 shows the results for the VG-LNA in the nominal 12dB gain state with Δf=50 MHz; Table 1 reports the results for all gain states and two-tone tests. Table 2 reports a summary of the performances and comparison with prior work [2-6]. The VG-LNA exhibits adequate performances compliant with the targeted switchless tapered phased-array TRX architecture, effective gain control, and record $P_C$ (1st sub-mW) and area on silicon. Despite the sub-mW $P_C$, the VG-LNA exhibits $S_{21}$ phase variation, $IP_{1dB}$, $BW_{3dB}$, and NF comparable with prior-art works. The sub-mW $P_C$ makes such a VG-LNA also suitable for mm-wave Internet of Things (IoT).

Table 1. Measured $IP_{1dB}$ and $IIP_3$ for all gain states and Δf = 50-400 MHz.

| $V_C$ [mV] | Gain [dB] 30GHz | $IP_{1dB}$ [dBm] | $IIP_3$ [dBm] | | | |
|---|---|---|---|---|---|---|
| | | | 50MHz | 100MHz | 200MHz | 400MHz |
| -740 | 8.0 | -21.0 | -12 | -11 | -9 | -6 |
| -530 | 9.0 | -22.3 | -13 | -12 | -10 | -8 |
| -310 | 10.0 | -23.5 | -14 | -13 | -12 | -11 |
| -92 | 11.0 | -24.5 | -15 | -14 | -13 | -13 |
| 152 | 12.0 | -25.5 | -16 | -15 | -15 | -14 |
| 403 | 13.0 | -26.3 | -17 | -16 | -16 | -16 |
| 683 | 14.0 | -27.2 | -18 | -17 | -17 | -17 |
| 1003 | 15.0 | -28.0 | -19 | -18 | -18 | -18 |
| 1340 | 16.0 | -28.8 | -19 | -19 | -19 | -19 |

Table 2. Comparison with prior-art CMOS VG-LNAs operating in 5G NR FR2.

| Ref. | Tech [nm] | $S_{21}$ [dB] | $f_C$ [GHz] | $BW_{3dB}$ [GHz] | NF [dB] | $IP_{1dB}$ [dBm] | Δφ$_{21}$* [°] | $P_C$ [mW] | Area [mm²] |
|---|---|---|---|---|---|---|---|---|---|
| **This** | 22 FDSOI | 16 / 8 | 30.1 / 30.1 | 7.1 / 7.2 | > 5.5 / > 6.3 | -28.8[3] / -21.0[3] | 16.7 | 0.97 / 0.41 | 0.044 |
| [2][1] | 22 FDSOI | 24.2 / 16.2[2] | 30.4[5] / 30.4[5] | 23.7 / 23.7[2] | > 2.4 / > 3.0 | -25 / -19 | - | 16 / 16 | - |
| [4] | 40 Bulk | 27.1 / 18.4 | 27.1 / 27.8 | 7.4 / 9.3 | > 3.3 / > 3.4 | -21.6 / -13.4 | 18 | 31.4 / 21.5 | 0.26 |
| [5] | 65 Bulk | 20.8 / 10.2 | 30.4[2] / 30.4[2] | 4 / 4[2] | > 3.7 / - | -20.4 / - | 8 | 26.7 / 16.5 | 0.20 |
| [3] | 90 Bulk | 21.4 / 11.6 | 37 / 37[2] | 11.3[4] / 11.3[2] | > 4.7 / - | -25.1[5] / -22[2] | 7.2 | 17.9 / 17.9 | 0.45[6] |
| [6] | 65 Bulk | 11.4 / -5 | 28 / 28 | 11.5 / - | > 4.7 / - | - / - | | 2.16 | 0.13 |

[1]Simulation results; [2]deduced from plots; [3]measured at 30 GHz; [4]dual band (26-30.5 GHz and 33.8-40.6 GHz); [5]calculated from available data; [6]including pads; *max. variation of ∠$S_{21}$ in $BW_{3dB}$.

## V. CONCLUSIONS

We have reported a 30GHz VG-LNA with 8dB gain-control range, record power consumption and area occupancy. Irrespective of the gain states, about 4-to-32 units of this VG-LNA consume less than one unit of the prior-art VG-LNAs.

ACKNOWLEDGMENTS

The authors are grateful to Keysight Technologies for their support through the donation of equipment and cad tools; Dr. C. Kretzschmar, Dr. P. Lengo, Dr. B. Chen (GlobalFoundries) for the technology support; Dr. D. Pepe (Renesas Design Zurich) for the technical discussions. This work was supported in part by the Poul Due Jensen Foundation, and in part by the European Commission through the European H2020 FET Open project IQubits (G.A. N. 829005).

## REFERENCES

[1] B. Sadhu et al., "A 28-GHz 32-Element TRX Phased-Array IC With Concurrent Dual-Polarized Operation and Orthogonal Phase and Gain Control for 5G Communications," *IEEE J. Solid-State Circuits*, vol. 52, no. 12, pp. 3373-3391, Dec. 2017.

[2] L. Gao and G. M. Rebeiz, "A 20-42-GHz IQ Receiver in 22-nm CMOS FD-SOI With 2.7-4.2-dB NF and -25-dBm IP1dB for Wideband 5G Systems," *IEEE Trans. Microw. Theory Techn.*, vol. 69, no. 11, pp. 4951-4960, Nov. 2021

[3] K. -C. Chang, Y. Wang and H. Wang, "A Broadband Variable Gain Low Noise Amplifier Covering 28/38 GHz bands with Low Phase Variation in 90-nm CMOS for 5G Communications," *IEEE MTT-S Int. Microw. Symp. (IMS)*, Atlanta, GA, USA, Jun. 2021, pp. 764-767.

[4] M. Elkholy, S. Shakib, J. Dunworth, V. Aparin and K. Entesari, "A Wideband Variable Gain LNA With High OIP3 for 5G Using 40-nm Bulk CMOS," *IEEE Microw. Wireless Compon. Lett.*, vol. 28, no. 1, pp. 64-66, Jan. 2018.

[5] S. Lee, J. Park and S. Hong, "A Ka-Band Phase-Compensated Variable-Gain CMOS Low-Noise Amplifier," *IEEE Microw. Wireless Compon. Lett.*, vol. 29, no. 2, pp. 131-133, Feb. 2019.

[6] H. Chen, H. Zhu, L. Wu, Q. Xue and W. Che, "A CMOS Low-Power Variable-Gain LNA Based on Triple Cascoded Common-Source Amplifiers and Forward-Body-Bias Technology," *IEEE MTT-S Int. Wireless Symp. (IWS)*, Rome, Italy, May 2021, pp. 1-3.

[7] J. J. Lee, "G/T and noise figure of active array antennas," *IEEE Trans. Antennas Propag.*, vol. 41, no. 2, pp. 241-244, Feb. 1993.

[8] M. Spasaro, F. Alimenti and D. Zito, "The Theory of Special Noise Invariants," *IEEE Trans. Circuits Syst. I: Reg. Papers*, vol. 66, no. 4, pp. 1305-1318, Apr. 2019.